\documentclass[prc,aps,showpacs,superscriptaddress,preprint,nofootinbib]{revtex4}

\usepackage{epsfig}
\usepackage[latin1]{inputenc}
\usepackage{float,amsmath,amssymb,slashed}
\usepackage{graphicx}

\newcommand{\beq}{\begin{equation}}
\newcommand{\eeq}{\end{equation}}
\newcommand{\bqa}{\begin{eqnarray}}
\newcommand{\eqa}{\end{eqnarray}}

\def\lsim{\mathrel{\rlap{\lower4pt\hbox{$\sim$}}
    \raise1pt\hbox{$<$}}}                
\def\gsim{\mathrel{\rlap{\lower4pt\hbox{$\sim$}}
    \raise1pt\hbox{$>$}}}                

\begin{document}

\title{Transverse momentum diffusion and collisional jet energy loss
  in non-Abelian plasmas} 
\author{Bj\"orn Schenke}
\affiliation{Department of Physics, McGill University, H3A\,2T8,
  Montreal, Quebec, Canada\vspace*{2mm}} \affiliation{Institut f\"ur
  Theoretische Physik, Goethe - Universit\"at Frankfurt
  \\ Max-von-Laue-Stra\ss{}e~1, D-60438 Frankfurt am Main,
  Germany\vspace*{2mm}} 
\author{Michael Strickland}
\affiliation{Institut f\"ur Theoretische Physik, Goethe -
  Universit\"at Frankfurt \\ Max-von-Laue-Stra\ss{}e~1, D-60438
  Frankfurt am Main, Germany\vspace*{2mm}} \affiliation{Gettysburg
  College, Gettysburg, PA 17325, USA\vspace*{2mm}} 
\author{\\Adrian Dumitru} 
\affiliation{Graduate School and University Center, City University of New York,
365 Fifth Avenue, New York, NY 10016, USA\vspace*{2mm}} 
\affiliation{RIKEN BNL Research Center, Brookhaven
  National Laboratory, Upton, NY 11973, USA\vspace*{2mm}}
\author{Yasushi Nara} \affiliation{Akita International University
  193-2 Okutsubakidai, Yuwa-Tsubakigawa, Akita-City, Akita 010-1211,
  Japan \\
  \vspace*{2mm}}
\author{Carsten Greiner}
\affiliation{Institut f\"ur Theoretische Physik, Goethe -
  Universit\"at Frankfurt \\
  Max-von-Laue-Stra\ss{}e~1,
  D-60438 Frankfurt am Main, Germany\vspace*{2mm}}

\begin{abstract}
  We consider momentum broadening and energy loss of high momentum
  partons in a hot non-Abelian plasma due to collisions.  We solve the
  coupled system of Wong-Yang-Mills equations on a lattice in real
  time, including binary hard elastic collisions among the
  partons. The collision kernel is constructed such that the total
  collisional energy loss and momentum broadening are lattice spacing
  independent. We find that the transport coefficient $\hat{q}$
  corresponding to transverse momentum broadening receives sizeable
  contributions from a power-law tail in the $p_\perp$-distribution of
  high-momentum partons. We establish the scaling of $\hat{q}$ and of
  $dE/dx$ with density, temperature and energy in the weak-coupling
  regime. We also estimate the nuclear modification factor $R_{AA}$
  due to elastic energy loss of a jet in a classical Yang-Mills field.
\end{abstract}
\maketitle

\section{Introduction}
The study of high transverse momentum jets produced in heavy-ion
collisions can provide information on the properties of the hot QCD
plasma produced in the central rapidity
region~\cite{Gyulassy:1993hr,Baier:1996sk,Baier:1998yf,Vitev:2002pf,Zakharov:2000iz,Jeon:2003gi,Salgado:2003gb,Wicks:2005gt,Jacobs:2005pk,Zhang:2007ja,Qin:2007rn,Fochler:2008ts}.
After the discovery of jet quenching at the Relativistic Heavy Ion
Collider (RHIC)~\cite{Adcox:2001jp,Adler:2002xw} a lot of progress has
been made towards using jets as a quantitative tomographic probe of
the QGP. Jet quenching refers to the suppression of high transverse
momentum hadrons, such as $\pi^0$ and $\eta$ mesons in central $Au+Au$
collisions compared to expectations from measurements in $p+p$
collisions. Whereas pions and $\eta$-mesons exhibit the same
suppression at high $p_\perp$, direct photons were found to be
unsuppressed~\cite{dirPhot_AuAu}. This indicates that the observed
suppression is related to the absorption (energy loss) of energetic
partons in the medium.

In this paper we study collisional energy loss and momentum broadening
of massless high momentum partons traversing a non-Abelian
plasma. Soft multi-particle interactions are treated by solving the
coupled system of Wong-Yang-Mills equations in real time. In addition,
particles can undergo hard elastic collisions.

So far, estimates based on perturbative QCD (pQCD) of the strength of
the coupling of jets to a plasma are sensitive to infrared
cutoffs. Also, they are often restricted to systems that are (at least
locally) in thermal equilibrium. The problem does not arise in the
Wong-Yang-Mills simulation
\cite{Wong:1970fu,Hu:1996sf,Krasnitz:1998ns,Krasnitz:1999wc,Krasnitz:2001qu,Krasnitz:2002mn,Krasnitz:2003jw,Krasnitz:2002ng,Dumitru:2005gp,Dumitru:2006pz},
since the soft sector is described by classical chromo-fields. It is
separated from the hard sector corresponding to hard elastic pQCD
processes. The soft sector is non-perturbative but in an essentially
classical way because the occupation number of field modes below the
saturation momentum (or temperature) are large
\cite{McLerran:1993ni,Mueller:2002gd}.

It is well known that a cutoff independent collisional energy loss can
be obtained by resumming soft
interactions~\cite{Braaten:1991jj,Braaten:1991we}. In the present
paper (see, also, Refs.~\cite{Dumitru:2007rp,Schenke:2008pm}) we show
by explicit implementation that this can also be achieved within the
framework of a transport theory by treating the soft interactions via
classical Yang-Mills fields defined on a lattice. We find that, in
practice, this works even for physical values of the gauge coupling,
$g\sim2$, so long as a weak-coupling (resp.\ continuum-limit)
condition specified in Eq.~(\ref{eq:continuum}) below is
satisfied. Within this framework, we are able to also consider the
interesting problem of elastic energy loss of a jet propagating
through a classical non-Abelian field, which might be relevant for
describing the early stages of a high-energy collision of large
nuclei; see below.

The main purpose of this paper is to extract lattice-spacing
independent results for the transport coefficient $\hat{q}$ associated
with broadening of the momentum distribution of gluon jets, as well as
for collisional energy loss $dE/dx$. Previous
publications~\cite{Dumitru:2007rp,Schenke:2008pm} already presented a
calculation of $\hat{q}$ within this approach, however lacking the
detailed analysis shown here as well as a computation of
$dE/dx$. Furthermore, in this paper we extract the entire $p_\perp^2$
distribution of jets passing through a thermal plasma (not only its
first moment $\hat{q}$). We also address the scaling of $\hat{q}$ and
$dE/dx$ with the particle density, temperature, and jet energy. The
scaling laws turn out to agree, qualitatively, with pQCD expectations
although the overall magnitude of $\hat{q}$ and $dE/dx$ is found to
receive substantial corrections.

We extract a value for $\hat{q}$ of $3.6\pm 0.3\,
\mathrm{GeV}^{2}\mathrm{fm}^{-1}$ at $T=400 \,\mathrm{MeV}$ in a
thermal SU(3) background for a parton with energy
$E=19.2\,\mathrm{GeV}$. For the collisional energy loss we obtain
$dE/dx = 1.6\pm 0.4$ \,GeV fm$^{-1}$.

This paper is organized as follows: We introduce the Boltzmann-Vlasov
equations as well as the Wong equations for non-Abelian plasmas in
Sec.~\ref{sec:transport}, and discuss the lattice implementation in
Sec.~\ref{sec:lattice}. We outline how collisions are included into
the Wong-Yang-Mills simulation in Sec.~\ref{sec:collisions}, and
describe how the separation between the soft and hard sector is done in
Sec.~\ref{sec:sepscale}. After discussing the initialization of the
simulation in Sec.~\ref{inenden}, we present results for collisional
energy loss and for momentum broadening in Sec.~\ref{sec:colleloss}. 
Finally, we close with conclusions in Sec.~\ref{sec:conc}.

\section{Boltzmann-Vlasov equation for non-Abelian gauge theories}
\label{sec:transport}
The classical transport theory for non-Abelian plasmas has been
established by Heinz and Elze
\cite{Heinz:1983nx,Heinz:1984yq,Heinz:1985qe,Elze:1989un}.  Here, we
solve numerically the classical transport equation for hard gluons
with adjoint SU(2) color charge $q=q^a \tau^a$, where the $\tau^a$ are
the color generators, including hard binary collisions
\begin{align}\label{eq:vlasov}
    p^\mu \left(\partial_\mu + g q^a F_{\mu\nu}^a \partial^\nu_{p} 
    + g f^{abc} A_\mu^b(x) q^c \partial_{q^a} \right)f={\cal C}\,.
\end{align}
$f=f(x,p,q)$ denotes the single-particle phase space distribution,
$F_{\mu\nu}^a=\partial_\mu A_\nu^a-\partial_\nu A^a_\mu-g
f^{abc}A_\mu^b A_\nu^c$ is the gauge field strength tensor, $g$ the
gauge coupling, $A_\mu^a$ the soft gauge field, and ${\cal C}$ is the
collision term to be defined below.  It is coupled self-consistently
to the Yang-Mills equation for the soft gluon fields,
\begin{equation}\label{eq:ym}
 D_\mu F^{\mu\nu} = j^\nu 
 = g \int \frac{d^3p}{(2\pi)^3}\,dq\,q\,v^\nu\,f(x,p,q)\,,
\end{equation}
with $v^\mu=(1,\mathbf{p}/p)$.  For ${\cal C}=0$ these equations
reproduce the ``hard thermal loop'' effective action near
equilibrium~\cite{Kelly:1994ig,Kelly:1994dh,Blaizot:1999xk}.  However,
the full classical transport theory~(\ref{eq:vlasov},\ref{eq:ym}) also
includes some higher $n$-point vertices of the dimensionally reduced
effective action for static gluons~\cite{Laine:2001my} beyond the
hard-loop approximation.  The back-reaction of the long-wavelength
fields on the hard particles (``bending'' of their trajectories) is
taken into account. This is essential for achieving cutoff independent
results for the transport coefficient $\hat{q}$ and for the energy
loss $dE/dx$ of high momentum partons.

\begin{figure}[htb]
  \begin{center}
    \includegraphics[width=12cm]{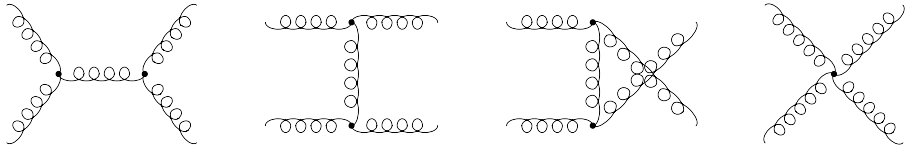}
    \caption{Processes contributing to $gg\rightarrow gg$ scattering
      at leading order.}
    \label{fig:ggline}
  \end{center}
\end{figure}

When the phase-space density is parametrically small, $f={\cal O}(1)$,
which is the case for hard momenta, the collision term is given by
\begin{align}
\label{eq:cterm1}
{\cal C} = \frac{1}{4E_1} \int_{\mathbf{p_2}}\,
\int_{\mathbf{p'_1}} \int_{\mathbf{p'_2}}(2\pi)^4\delta^{(4)}
(p'_1+p'_2-p_1-p_2)\left(f'_1 f'_2 |{\cal M}_{1'2'\to 12} |^2 -  f_1 f_2
|{\cal M}_{12\to 1'2'}|^2 \right) \,,
\end{align}
with $\int_{\mathbf{p_i}}=\int \frac{d^3p_i}{(2\pi)^3 2E_i}$.  The
matrix element ${\cal M}$ includes all $gg\rightarrow gg$ tree-level
diagrams shown in Fig. \ref{fig:ggline}, and color factors as
appropriate for the SU(2) gauge group.

We employ the test particle method and replace the
continuous distribution $f(x,p,q)$ by a large number of test
particles \cite{Bertsch:1988ik}:
\begin{equation}
    f(\mathbf{x},\mathbf{p},q)=\frac{1}{N_{\text{test}}}\sum_i \delta^3(\mathbf{x}-\mathbf{x}_i(t))(2\pi)^3\delta^{(3)}(\mathbf{p}-\mathbf{p}_i(t))\delta^{(N^2-1)}(q-q_i(t))\,,
\end{equation}
which leads to the Wong equations~\cite{Wong:1970fu} (also see \cite{Litim:1999id,Litim:1999ns})
\begin{align}
    \dot{\mathbf{x}}_i(t)&=\mathbf{v}_i(t)\,, \label{wong1}\\
    \dot{\mathbf{p}}_i(t)&=g
    q^a_i(t)\left(\mathbf{E}^a(t)+\mathbf{v}_i(t)\times\mathbf{B}^a(t)\right)
    \,,\label{wong2}\\ 
    \dot{q}_i(t)&=-igv_i^\mu(t)[A_\mu(t),q_i(t)]\,.\label{wong3}
\end{align}
Here, $\mathbf{x}_i(t)$, $\mathbf{v}_i(t)$, and $q_i(t)$, are the
position, velocity\footnote{We consider only massless particles here
  so that $|\mathbf{v}_i|=1$.} and color charge of the $i^{\rm th}$ test
particle. $N_{\mathrm{test}}$ denotes the number of test particles per
physical particle, $A_\mu(t)=A_\mu^a(t) \tau^a$ and the commutator
commutes color generators $\tau^a$.  The last equation (\ref{wong3})
describes the precession of the color charge $q_i(t)$ due to the color
fields.

Writing the current in terms of the individual test particles, the
Yang-Mills equation for the soft gluon fields becomes
\begin{equation}\label{currentwym}
    D_\mu F^{\mu\nu}=J^\nu=\frac{g}{N_{\mathrm{test}}}\sum_i q_i v_i^\nu\delta(\mathbf{x}-\mathbf{x}_i(t))\,.
\end{equation}

The theory without collisions as given by
equations~(\ref{wong1}-\ref{wong3}) coupled to the lattice Yang-Mills
equations (\ref{currentwym}) was first solved in \cite{Moore:1997sn}
to study Chern-Simons number diffusion in non-Abelian gauge theories
at finite temperature.  It was applied later also to the problem of
gauge-field instabilities in anisotropic SU(2)
plasmas~\cite{Dumitru:2005gp,Dumitru:2006pz}.  Our numerical
implementation is based on the improved formulation detailed in
\cite{Dumitru:2006pz} where the non-Abelian currents, generated by the
hard particle modes on the lattice sites, are ``smeared''. This
technique makes simulations in three dimensions on large lattices
possible in practice.

\section{Real-time lattice simulation}
\label{sec:lattice}
The time evolution of the Yang-Mills field is determined by the
standard Hamiltonian method in $A^0=0$ gauge
\cite{Ambjorn:1990pu,Hu:1996sf,Moore:1997sn}. The temporal gauge is
particularly useful because it allows for a simple identification of
the canonical momentum as the electric field
\begin{equation}
    \mathbf{E}^a=-\dot{\mathbf{A}}^a\,.
\end{equation}
In addition, time-like link variables $U$, defined below, become simple
identity matrices.

The lattice Hamiltonian in this gauge is given by \cite{Kogut:1974ag}
\begin{equation}\label{KSH}
    H_L=\frac{1}{2}\sum_i \mathbf{E}_{L\,i}^{a\,2}+\frac{1}{2}\sum_\square\left(N_c-\text{Re}\text{Tr}U_\square\right)+\frac{1}{N_{\text{test},\,L}}\sum_j|\mathbf{p}_{L\,j}|\,,
\end{equation}
including the particle contribution
$1/N_{\text{test},\,L}\sum_j|\mathbf{p}_{L\,j}|$.  The plaquette is
defined by
\begin{equation}
    U_\square=U_x(i)U_y(i+\hat{x})U_x^\dag(i+\hat{y})U_y^\dag(i)\,,
\end{equation}
with the link variable
\begin{equation}\label{linkdef}
    U_\mu(i)=e^{iagA_\mu(i)}\,.
\end{equation}
Note that the index $\mu$ on $U$ is merely an indicator of its direction and not a Lorentz index.
The shifts $\hat{x}$ and $\hat{y}$ are one lattice spacing in length
and directed into the $x$- or $y$- direction, respectively.
\footnote{We set $\tau_a=\sigma_a$, the Pauli matrices, without the
  usual factor of $1/2$, i.e., the commutation relation reads
  $[\tau^a,\tau^b]=2\delta^{ab}$. Another factor of $1/2$ is absorbed
  into the $A$-field, which has to be taken into account when
  calculating the physical fields $\mathbf{E}$ and $\mathbf{B}$ from
  it.}

Eq. (\ref{KSH}) is given in lattice units, which are chosen such that
all lattice variables are dimensionless:
\begin{align}\label{latticevariables}
    \mathbf{E}^a_L=\frac{ga^2}{2}\mathbf{E}^a\,, && \mathbf{B}^a_L=\frac{ga^2}{2}\mathbf{B}^a\,, && \mathbf{p}_L=\frac{a}{4}\mathbf{p}\,, && Q_L^a=\frac{1}{2}q^a\,, && N_{\text{test},\,L}=\frac{1}{g^2}N_{\text{test}}\,,
\end{align}
with the lattice spacing $a$. $H_L$ is hence related to the
physical Hamiltonian by $H=4/(g^2a)\, H_L$. 
To convert lattice variables to physical units we
will fix the lattice length $L$ in fm, which will then determine
the physical scale for $a$. All other dimensionful quantities can
then be determined from Eqs.~(\ref{latticevariables}). The
Hamiltonian (\ref{KSH}) determines the energy density of the
system and enters the equations of motion for the
fields, e.g.,
\begin{equation}
    \frac{d}{dt}\mathbf{E}_L=\left\{\mathbf{E}_L,H_L\right\}\,,
\end{equation}
with the Poisson bracket $\{\cdot,\cdot\}$.
Our lattice has periodic boundary conditions in all spatial directions.

\section{Collisions}
\label{sec:collisions}
The collision kernel (\ref{eq:cterm1}) is similar to that used in
(parton) cascade simulations
\cite{PhysRevC.29.2146,Geiger:1991nj,Kortemeyer:1995di,Zhang:1998tj,Cheng:2001dz,PhysRevC.40.2611,Molnar:2001ux,Xu:2004gw,Xu:2004mz,Xu:2007aa}.
Here, it is restricted to hard binary collisions since soft
multi-parton interactions are mediated by interactions with the
collective Yang-Mills field. This way, we are able also to study
collective phenomena and their contribution to isotropization and
thermalization. In particular, we can in principle also study systems
away from equilibrium (see \cite{Dumitru:2007rp}) for which the scale
corresponding to the Debye-mass squared in an isotropic system becomes
negative~\cite{Mrowczynski:1996vh,Romatschke:2003ms,Mrowczynski:2004kv,Romatschke:2004jh}. In
this case it can obviously not damp the propagator to act as a cut off
for the momentum exchange in the infrared.

To complete our dual particle/field description, we need to specify
the separation scale ${k^*}$ between the field and particle degrees of
freedom. We will discuss this separation scale in detail below. For
now it will serve as a lower bound for the exchanged momenta for
binary elastic particle collisions. All softer momentum exchanges are
mediated by the fields.

The collision term (\ref{eq:cterm1}) is incorporated using the
stochastic method introduced and applied in
\cite{Danielewicz:1991dh,Lang:1993dh,Xu:2004mz}.  We do not interpret
the cross section in a geometrical way as done in
\cite{PhysRevC.29.2146,Kortemeyer:1995di,Zhang:1998tj,Cheng:2001dz,PhysRevC.40.2611,Molnar:2001ux}
but determine scattering processes in a stochastic manner by sampling
possible transitions in a volume element per time interval.  This
collision algorithm can be extended to include inelastic processes
$gg\leftrightarrow ggg$ as done in \cite{Xu:2004mz,Xu:2007aa}, which
will also be incorporated in the future in our simulations.

The collision rate in a spatial volume element $\Delta^3 x$ per unit
phase space for a particle pair with momenta in the range (${\bf p}_1,
{\bf p}_1+\Delta^3 p_1$) and (${\bf p}_2, {\bf p}_2+\Delta^3 p_2$)
follows from Eq.~(\ref{eq:cterm1})
\begin{align}
\label{collrate22}
\frac{\Delta N_{\text{coll}}}{\Delta t \frac{1}{(2\pi)^3} \Delta^3 x
\Delta^3 p_1} &= \frac{1}{2E_1}
\frac{\Delta^3 p_2}{(2\pi)^3 2E_2} f_1 f_2 \nonumber \\
& \times \frac{1}{2} \int \frac{d^3 p^{'}_1}{(2\pi)^3 2E^{'}_1}
\frac{d^3 p^{'}_2}{(2\pi)^3 2E^{'}_2} | {\cal M}_{12\to 1'2'} |^2 (2\pi)^4
\delta^{(4)} (p_1+p_2-p^{'}_1-p^{'}_2).
\end{align}
Expressing the distribution functions as
\begin{equation}
\label{distf}
f_i=\frac{\Delta N_i}{\frac{1}{(2\pi)^3} \Delta^3 x \Delta^3 p_i},
\quad i=1,2\,,
\end{equation}
and employing the usual definition of the cross section
for massless particles \cite{Groot:1980}
\begin{equation}
\label{cs22}
\sigma_{22}= \frac{1}{4s} \int
\frac{d^3 p^{'}_1}{(2\pi)^3 2E^{'}_1} \frac{d^3 p^{'}_2}{(2\pi)^3 2E^{'}_2}
| {\cal M}_{12\to 1'2'} |^2 (2\pi)^4 \delta^{(4)} (p_1+p_2-p^{'}_1-p^{'}_2)
\,,
\end{equation}
one obtains the total collision probability in a volume element $\Delta^3 x$
and time interval $\Delta t$:
\begin{equation}
\label{p22}
P_{22} = \frac{\Delta N_{\text{coll}}}{\Delta N_1 \Delta N_2} =
\tilde{v}_{\text{rel}} \sigma_{22} \frac{\Delta t}{\Delta^3 x}\,.
\end{equation}
$\tilde{v}_{\text{rel}}=s/2E_1E_2$
denotes the relative velocity, where $s$ is the invariant
mass of the particle pair. $P_{22}$ is a number between $0$
and $1$.\footnote{ In practice one has to choose suitable
$\Delta^3 x$ and $\Delta t$ to ensure that $P_{22}<1$.}
Whether or not a collision occurs is sampled stochastically as
follows: We compare $P_{22}$ to a uniformly distributed random
number between $0$ and $1$. If the random number is less than
$P_{22}$, the collision does occur. Otherwise, there is no
collision between the two particles in that time step.
Since we represent each physical particle by $N_{\text{test}}$ test particles, we
have to rescale the cross section as $\sigma\rightarrow
\sigma/N_{\text{test}}$. This leads to
\begin{equation}
\label{p22test}
P_{22} = \tilde{v}_{\text{rel}} \frac{\sigma_{22}}{N_{\text{test}}}
\frac{\Delta t}{\Delta^3 x}\,.
\end{equation}
To determine this probability, we require the total cross section
$\sigma_{22}$. To leading order in $\alpha_s$ it follows from the
differential cross section obtained from the diagrams in
Fig.~\ref{fig:ggline}
\cite{Combridge:1977dm,Owens:1977sj,Bern:2002tk}:
\begin{equation}\label{eq:xsec}
    \frac{d\sigma}{dt}=\frac{4\pi\alpha_s^2}{s^2} \frac{N_c^2}{N_c^2-1}\left(3-\frac{tu}{s^2}-\frac{su}{t^2}-\frac{st}{u^2}\right)\,,
\end{equation}
with $N_c$ the number of colors.
The invariant Mandelstam variables are
\begin{align}
    &s = (p_1+p_2)^2\,, ~t = (p_1-p_1')^2\,, ~u = (p_1-p_2')^2\,.
\end{align}
Using $t=-{q}^2$, with $q$ the momentum transfer, and the identity
$s+t+u=0$ for massless particles, we can express the total cross
section for processes with $\sqrt{q^2}$ larger than $k^*$ as
\begin{equation}\label{totcross}
    \sigma_{22}=\int_{{k^*}^2}^{s/2}\frac{d\sigma}{dq^2}dq^2\,.
\end{equation}
The momentum transfer is then determined stochastically in the
center-of-momentum frame of the two colliding particles from the
probability distribution
\begin{equation}
 {\cal P}(q^2)=\frac{1}{\sigma_{22}}\, \frac{d\sigma}{dq^2}\,.
\end{equation}
In Eq.~(\ref{totcross}) we have introduced the cutoff $k^*$ for
point-like binary collisions. To avoid double-counting, this cutoff
should be on the order of the hardest field mode that can be
represented on the given lattice, $k^*\simeq\pi/a$.

\section{Separation scale}
\label{sec:sepscale}
The scattering processes in the regime of hard momentum exchange are
described by elastic binary collisions, while soft momentum exchanges
are mediated by the fields. A scattering in the soft regime
corresponds to deflection of a particle in the field of the other(s).

Physically, the separation scale ${k^*}$ should be sufficiently small
so that the soft field modes below ${k^*}$ are highly
occupied~\cite{Ambjorn:1990pu} and hence can be described
classically. On the other hand, ${k^*}$ should be sufficiently large
to ensure that hard modes can be represented by particles and that
collisions are described by~(\ref{eq:cterm1}), which is valid only for
low occupation numbers since the Bose enhancement factor $(1+f)$ is
approximated by 1. In practice $g\sim 1$, and we choose ${k^*}$ to be
on the order of the temperature.  At the same time, ${k^*}$ is related
to the hardest available field mode, which on a cubic lattice is given
by $\sqrt{3}\pi/a$. Obviously, any matching between soft and hard
regimes can only be done approximately, because the lattice on which
the field modes are defined is cubic, while the momentum space cutoff
of the hard collision integral is implicitly spherical.

The ``soft'' scale is given by
\begin{equation}\label{minfty}
  m^2_{D} = \frac{2g^2N_c}{N_c^2-1} 
 \int\frac{d^3p}{(2\pi)^3} \frac{f(\mathbf{p})}{|\mathbf{p}|}
             \sim \frac{\pi^2}{2}\, 
                 \frac{g^2 N_c}{N_c^2-1} \, \frac{n}{p_h}\,,
\end{equation}
where $N_c=2$ is the number of colors and $n$ denotes the number
density of hard gluons, summed over two helicities and $N_c^2-1$
colors. Also, $p_{\text{h}}\approx 3T$ is the typical momentum of
a hard particle from the medium.

To allow for reliable numerical simulations one should have $m_D L\gg
\pi$ and $m_D a\ll \pi$. The first condition ensures that the relevant
soft modes actually fit on the lattice while the latter ensures that
the lattice can resolve the wavelength $1/m_D$ to good precision.

As we have argued above, we choose the inverse lattice spacing
to be on the order of the temperature of the medium. Thus,
with~(\ref{minfty}) the condition $m_D a \ll \pi$ roughly
translates to
\begin{equation} \label{eq:continuum}
\frac{g^2 N_c}{N_c^2-1} \, \frac{n}{T^3} \ll 1~.
\end{equation}
In order to satisfy this relation, which is essentially the
weak-coupling condition, at $g\sim1$, we perform the numerical
simulations below for an extremely hot and undersaturated medium:
$T^3\gg n$. This ensures that the simulations are carried out near the
continuum limit. We verify below that transverse momentum
broadening of a high-energy jet passing through a thermal medium is
independent of $T$ if the density and the ratio of jet momentum to
temperature is fixed. One may therefore obtain a useful
weak-coupling estimate of $\langle p_\perp^2\rangle$ (resp.\ for
the related transport coefficient $\hat{q}$) by extrapolating our
measurements down to temperatures relevant to present heavy-ion
collisions.

\section{Initialization}
\label{inenden}
We consider a heat-bath of Boltzmann
distributed particles with a density of $n=\{5,10,20\}\,\mathrm{fm}^{-3}$ and
an average particle momentum of $3T=\{6,12,18,24\}$~GeV.
For a given lattice (resp.\ ${k^*}$) we take the initial
energy density of the thermalized fields to be
\begin{equation}\label{epsfields}
\varepsilon_\text{fields}=\int \frac{d^3k}{(2\pi)^3} \,k \hat{f}_{\rm Bose}(k)\Theta(k^*-k)\,,
\end{equation}
where 
\begin{equation}
\hat{f}_{\rm Bose}(k)=\frac{n\, \pi^2}{T^3 \zeta(3)}\,\frac{1}{e^{k/T}-1}
\end{equation}
is a Bose distribution normalized to the assumed particle density
$n$, and $\zeta$ is the Riemann zeta function.

The initial field amplitudes are sampled from a Gaussian distribution:
$ \langle
A_i^a(x)A_j^b(y)\rangle = \frac{4\mu^2}{g^2} \delta_{ij} \delta^{ab}
\delta(\mathbf{x}-\mathbf{y})\,.$
To thermalize the initial fields (approximately), we match their
Fourier spectrum to the classical limit of the Bose
distribution. Hence, the initial spectrum is gauge-fixed to Coulomb
gauge and a filter is applied such that
$$A_i\sim 1/k$$ (in continuum notation).  
Setting $E_i=0$
initially\footnote{Equipartitioning of electric and magnetic fields is
  achieved very rapidly within a few time steps.}, Gauss's law implies
that the local charge density at time $t = 0$ vanishes. We ensure that
any particular initial condition satisfies exact local charge
neutrality. The charge smearing algorithm for SU(2) explicitly
exploits (covariant) current conservation and hence Gauss's law is
satisfied exactly by construction at all times~\cite{Dumitru:2006pz}.

\begin{figure}[hbt]
  \begin{center}
    \includegraphics[width=11cm]{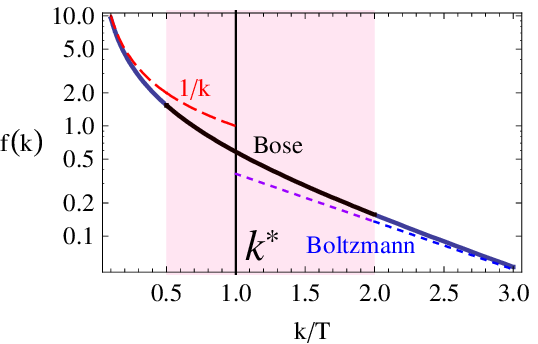}
    \caption{(Color online) Bose distribution and its low and
      high-momentum limits,
      used for the initial fields and particles,
      respectively. Physically, the separation ${k^*}$ should be on
      the order of the temperature $T$. The band between $T/2$ and
      $2T$ roughly indicates the region within which we vary
      $k^*$.}
    \label{fig:distributions}
  \end{center}
\end{figure}

The above procedure ensures that there is no large discontinuity of the
energy density when going from the field to the particle regime.
This way we are able to vary the separation scale ${k^*}$ about
the temperature $T$ by varying the lattice spacing. Fig.~\ref{fig:distributions}
shows the distribution of field modes and particles and the separation
scale ${k^*} \sim T$.

\section{Momentum diffusion and energy loss of high momentum partons}

Having initialized the background particles and fields, we can now add
a few high-momentum test particles propagating along a given
(``longitudinal'') direction which represent the jets. Their density
should be sufficiently low so that they do not influence the thermal
background significantly and so their mutual interaction is
minimized.

We always initialize ``bunches'' of test particles, which represent
one physical hard momentum parton (``jet'').  A bunch corresponds to
$N_{\mathrm{test}}$ particles in the same lattice cell. The physical
color charge is independent of $N_{\mathrm{test}}$. If, in fact, the
color charges of all test particles representing one jet add to zero,
no coherent radiation is emitted (colorless jet). Such jets can only
suffer collisional energy loss\footnote{Note that individual test
  particles from the bunch are of course {\em colored} and hence they
  collide not only with hard thermal particles but also with the modes
  of the thermal fields.}.  In the particle-in-cell simulation
radiative energy loss is not consistenly included (see
e.g. \cite{Hededal:2005hz}). Initializing a bunch of test particles
with aligned color vectors, leading to a net current on the lattice,
will hence not correspond to the correct physical bremsstrahlung
process.  We postpone the consistent implementation of radiative
energy loss to future work.

\begin{figure}[htb]
  \begin{center}
    \includegraphics[width=11cm]{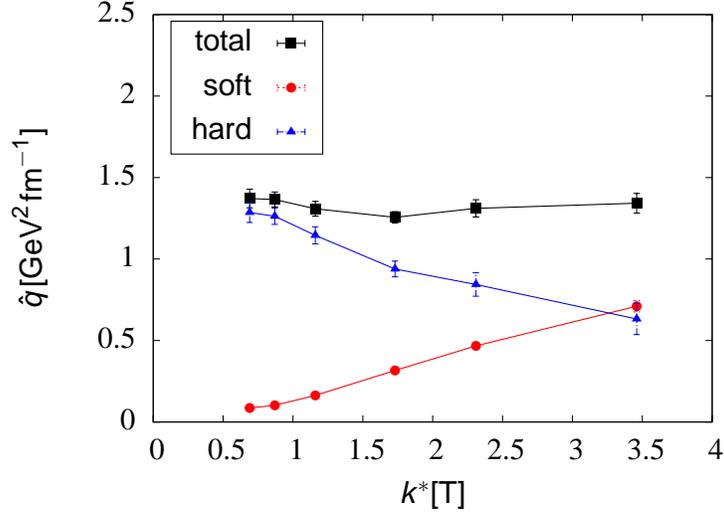}
    \caption{(Color online) (In-)dependence of the transport coefficient $\hat{q}$ on
      the separation scale $k^*$. $T=4$ GeV, $g=2$,
      $n=5\,\mathrm{fm}^{-3}$.}
    \label{fig:qhatk}
  \end{center}
\end{figure}

\label{sec:colleloss}
As detailed above, colorless bunches of test particles permit us to
restrict to collisional energy loss and momentum broadening due to
elastic collisions only. We first demonstrate that in our approach
both
\begin{equation}\label{eq:qhat}
 	\hat{q}=\frac{1}{\lambda \sigma} \int
        d^2p_\perp\,p_\perp^2\frac{d\sigma}{d p_\perp^2}\,,
\end{equation}
and the differential energy loss $dE/dx$ are independent of the
separation scale $k^*$. Here and in what follows, $p_\perp$ denotes
the momentum transverse to the initial jet momentum. $\hat{q}$ can be
extracted from the squared transverse momentum of the test-particles 
accumulated up to a time $t$:
\begin{equation}
 	\hat{q}=\frac{\langle p_\perp^2\rangle(t)}{t}\,.
\end{equation}
Fig.~\ref{fig:qhatk} depicts the contributions to $\hat{q}$ due to
soft and hard collisions, respectively, as well as the total. In these
simulations the gluon density of the medium was taken to be
$n=5\,\mathrm{fm}^{-3}$, the temperature $T=4\,\mathrm{GeV}$, and
the jet energy is $16$ times the average thermal momentum
($48\,T$). $\hat{q}$ is shown as a function of the separation scale
$k^*\sim \sqrt{3}\pi/a$.

The curves for the total and the soft contributions were averaged over
$\sim 80$ runs for each point. The curve corresponding to the hard
sector was obtained by subtracting the result without hard collisions
(soft sector only) from the total.  The error bars indicate one
standard deviation about the mean.

We find that the total value is constant to a good approximation
although the contribution due to soft scatterings changes
considerably. Below $k^*\approx T$, the soft sector contributes less
than $10\%$, while it starts dominating around $k^*\approx 3T$. It is
evident, therefore, that transport coefficients obtained in the
leading logarithmic (LL) approximation from the pure Boltzmann
approach (without soft fields) are rather sensitive to the infrared
cutoff $k^*$, unless the energy $\sqrt{s}$ is extremely high. In LL
approximation,
\begin{equation}    \label{<kt2>LL}
  \hat{q}  = n \, \frac{4 \pi\, \alpha_s^2 N_c^2}{N_c^2-1}
    \ln\left(C^2\frac{Q^2}{k^{* 2}} \right) \,,
\end{equation}
where $Q^2\simeq s$ is the upper bound for the momentum transfer and
$C$ is a constant.

\begin{figure}[htb]
  \begin{center}
    \includegraphics[width=11cm]{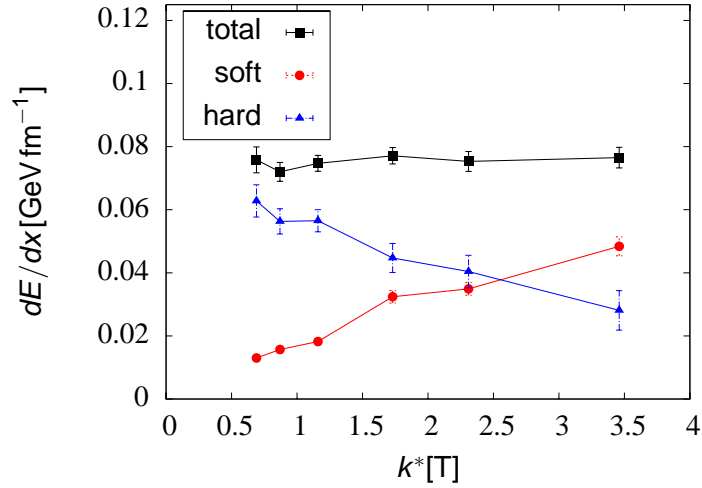}
    \caption{(Color online) (In-)dependence of the energy loss $dE/dx$ on the
      separation scale $k^*$. $T=4$ GeV, $g=2$,
      $n=5\,\mathrm{fm}^{-3}$.}
    \label{fig:dEdxk}
  \end{center}
\end{figure}

Fig.~\ref{fig:dEdxk} repeats the same analysis for the collisional
energy loss per unit path length, $dE/dx$. Again we find a constant
total energy loss, and a similar dependence on $k^*$ of the partial
contribution due to soft interactions as for $\hat{q}$.
We have also verified the $k^*$-independence for different
temperatures, densities and jet energies. Thus, the above-mentioned
matching of soft and hard processes provides estimates for $\hat{q}$
and $dE/dx$ which are independent of the artificial separation scale
$k^*$ (and of the lattice spacing $a$). It cures the infrared
divergence of the perturbative hard-scattering cross section and does
not rely on infrared cutoffs from equilibrium physics such as the
Debye mass; thus, calculations are not restricted to equilibrium. On
the other hand, the matching procedure might have to be modified for
other observables which are sensitive to very different scales. This
should be analyzed in the future.

\begin{figure}[htb]
  \begin{center}
    \includegraphics[width=11cm]{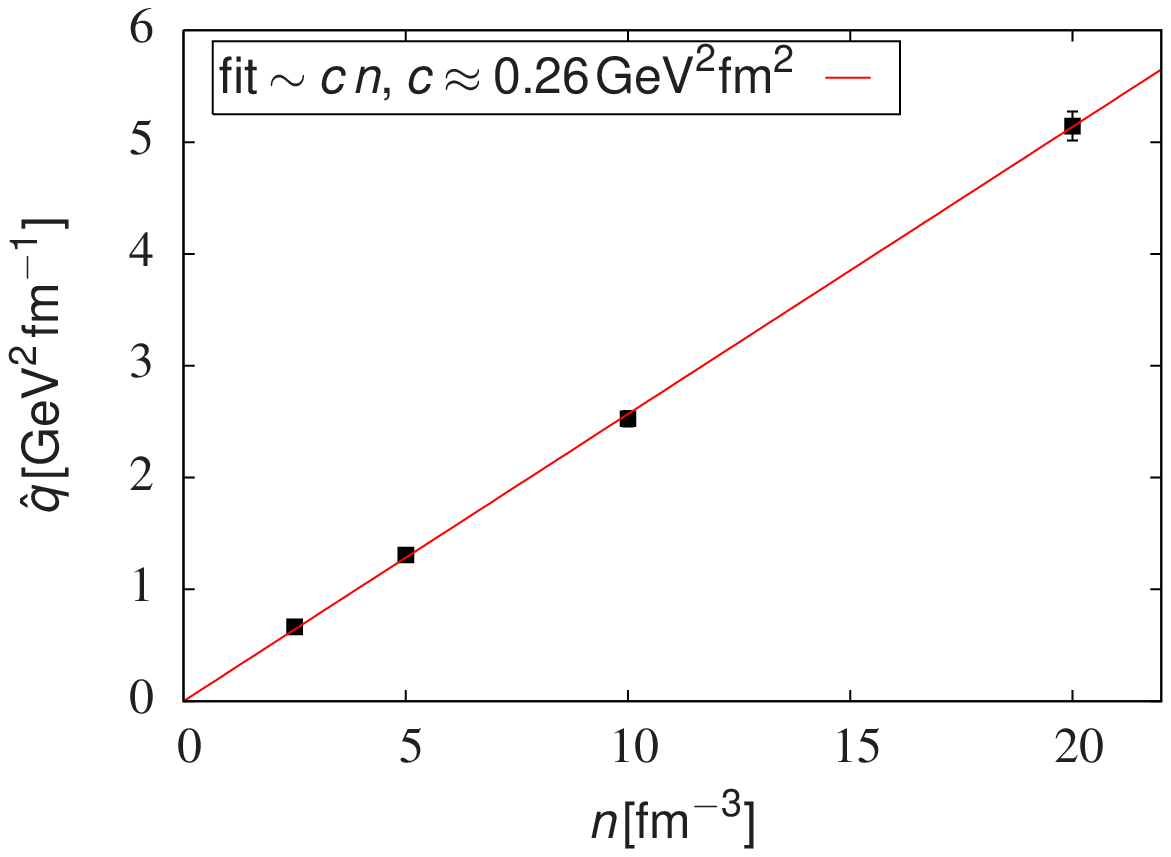}
    \caption{(Color online) Linear dependence of $\hat{q}$ on the density $n$. $T=4$
      GeV, $g=2$, $E/T=48$, $k^*\approx 1.16\,T$. The line shows the
      best linear fit.}
    \label{fig:qhatn}
  \end{center}
\end{figure}

\begin{figure}[htb]
  \begin{center}
    \includegraphics[width=11cm]{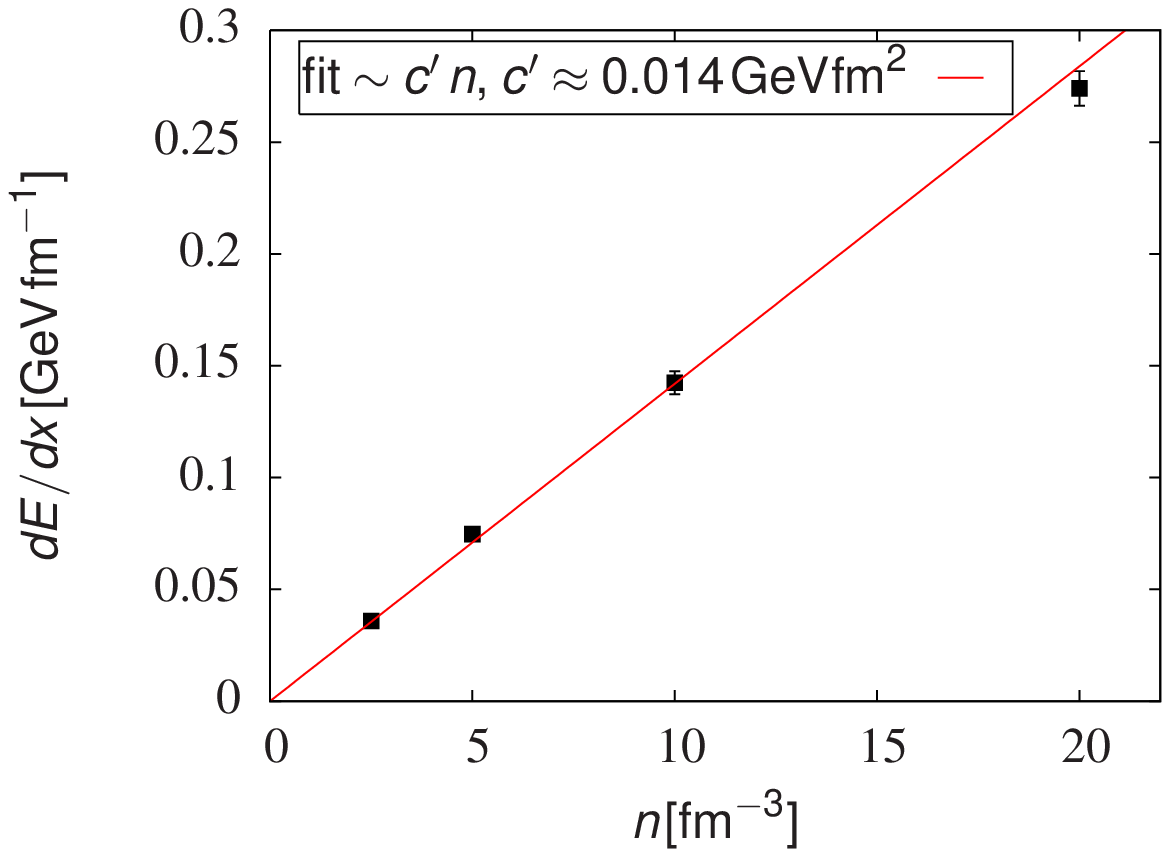}
    \caption{(Color online) Linear dependence of $dE/dx$ on the density $n$. $T=4$
      GeV, $g=2$, $E/T=48$, $k^*\approx 1.16\,T$. The line shows the
      best linear fit.}
    \label{fig:dEdxn}
  \end{center}
\end{figure}

\begin{figure}[htb]
  \begin{center}
    \includegraphics[width=11cm]{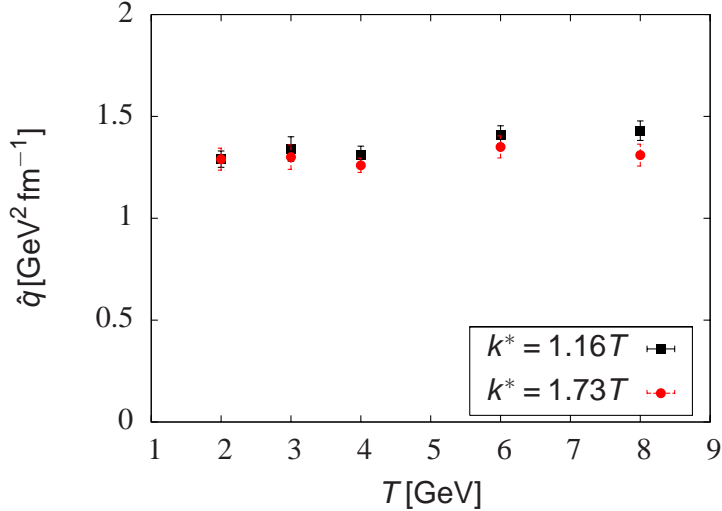}
    \caption{(Color online) (In-)dependence of $\hat{q}$ on the temperature
      $T$. $n=5$ fm$^{-3}$, $g=2$, $E/T=48$, $k^*\approx 1.16\,T$
      (squares) and $k^*\approx 1.73\,T$ (circles).}
    \label{fig:qhatT}
  \end{center}
\end{figure}

\begin{figure}[htb]
  \begin{center}
    \includegraphics[width=11cm]{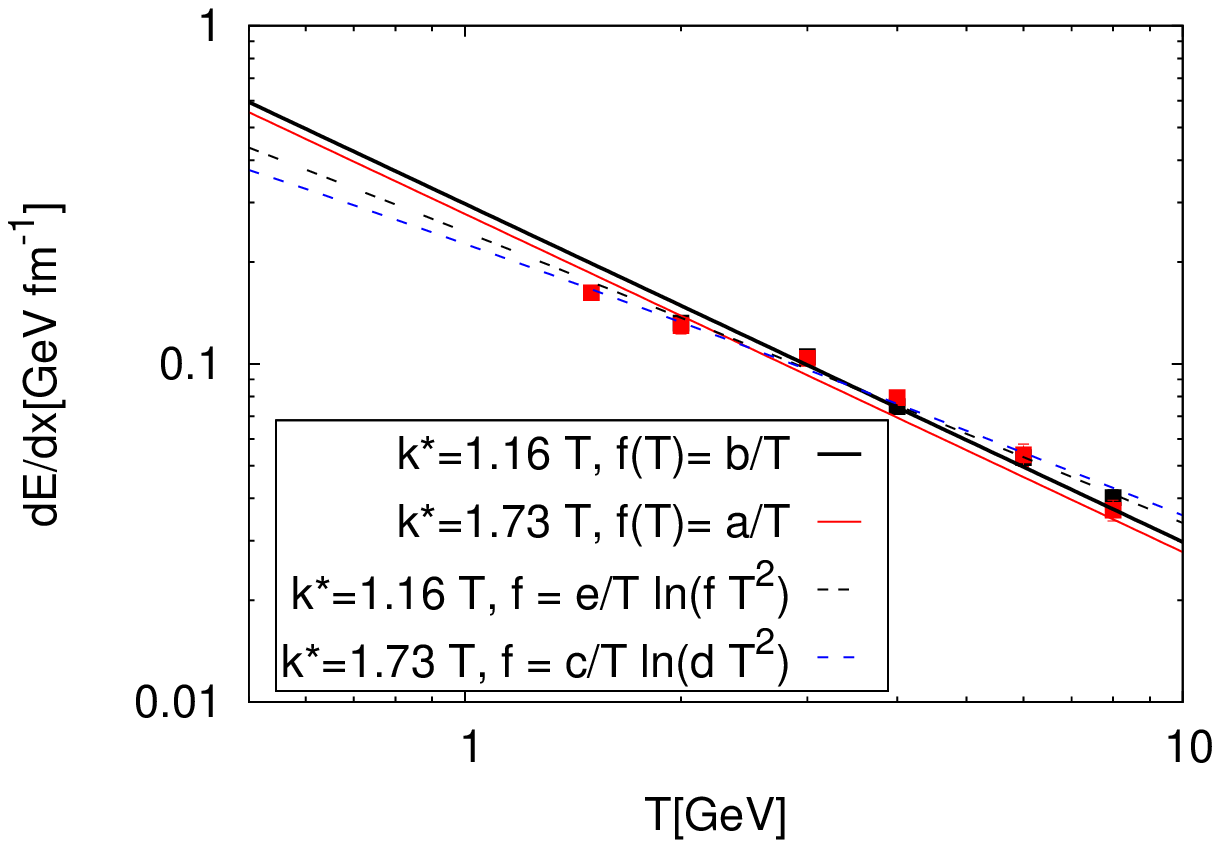}
    \caption{(Color online) Dependence of $dE/dx$ on the temperature $T$. $n=5$
      fm$^{-3}$, $g=2$, $k^*\approx 1.16\,T$ (black squares) and
      $k^*\approx 1.73\,T$ (red/grey squares). $a \dots f$ are fit
      parameters - the plot shows the best fits. Extrapolation to
      temperatures $T\sim 500\,\mathrm{MeV}$ leads to uncertainties of
      order 40 \%.}
    \label{fig:dEdxT}
  \end{center}
\end{figure}

Next, we turn to the density and temperature dependence of $\hat{q}$
and $dE/dx$. Figs.~\ref{fig:qhatn} and \ref{fig:dEdxn} show the
linear rise of $\hat{q}$ and $dE/dx$ with the density, which is
expected from Eq.~(\ref{eq:qhat}) and the perturbative results
(\ref{<kt2>LL}) and (\ref{eq:dEdx}) below, respectively. We will use
this linear dependence below to extrapolate to larger densities
(e.g., $n^{\mathrm{thermal}}(T=500\, \mathrm{MeV})\approx 32 \,
\mathrm{fm}^{-3}$ for pure glue in SU(3)).

Fig.~\ref{fig:qhatT} shows that $\hat{q}$ is approximately independent
of $T$ as long as the ratio of the jet energy to the temperature $E/T$
as well as the density $n$ are fixed. From (\ref{<kt2>LL}) we expect
at most a logarithmic dependence on $T$, because $Q^2\simeq s$ and
$\langle s \rangle=6 E T$. The simulation shows that this dependence
is very weak.

Fig.~\ref{fig:dEdxT} shows $dE/dx$ dropping approximately like $\sim
1/T$. This behavior is expected from the perturbative LL result for
elastic energy loss (see, for example, ref.~\cite{Dumitru:2000up})
\begin{equation}\label{eq:dEdx}
 	\frac{dE}{dx}=n\left(\frac{16\pi\, \alpha_s^2
          N_c^2}{N_c^2-1}\right)
        \frac{E}{s}\ln\left(C'^2\frac{Q^2}{{k^*}^2}\right)\,,
\end{equation}
where $s$ is the center-of-mass energy for a process involving
scattering of the jet from a hard thermal excitation, $Q^2\simeq s$ is
the upper limit for the momentum transfer, and $C'$ is a
constant. Because $\langle s \rangle=6 E T$ this leads to $dE/dx\sim
1/T$. Additionally, $T$ also appears in the logarithm, but this
dependence turns out to be weak.

We can now extrapolate to temperatures which are accessible in
practice, for which direct computations can not be performed due to
the numerical reasons explained above. For an ideal gas of thermal
gluons at a temperature $T$ the density $n=16 T^3 \zeta(3)/\pi^2$
(for $N_c=3$). Using the linear dependence of $\hat{q}$ on $n$
confirmed above, and its independence on $T$ for fixed $n$ and
$E/T$, we find $\hat{q}\approx 7\pm
0.6~\mathrm{GeV}^{2}\mathrm{fm}^{-1}$ at $T=500\, \mathrm{MeV}$. This
number has been rescaled to the color factors appropriate to
SU(3) \footnote{We divide the results by the prefactors given in
  Eqs.~(\ref{<kt2>LL}) and (\ref{eq:dEdx}), respectively, which
  correspond to $N_c=2$, and multiply by the prefactors appropriate
  for $N_c=3$.}. Since $E/T$ is fixed, this result corresponds to a
jet energy of $E=48\,T=24\,\mathrm{GeV}$. The quoted error arises from
using different possible fits, including a logarithmic dependence on
$T$ or not, and from different choices for $k^*$. For a temperature of
$T=400\, \mathrm{MeV}$ and the corresponding thermal gluon density, we
find $\hat{q}\approx 3.6\pm 0.3\,\mathrm{GeV}^{2}\mathrm{fm}^{-1}$
($E=19.2\,\mathrm{GeV}$). We emphasize that our simulations do not
account for quarks and anti-quarks which would provide a sizeable
contribution to the thermal density. Nevertheless, such values for
$\hat{q}$ are within the range extracted from RHIC
data~\cite{Baier:2006fr,Majumder:2007iu,Bass:2008rv}.

In Fig.~\ref{fig:dEdxT} we present a possible extrapolation of $dE/dx$
to temperatures around $500$~MeV. We find $dE/dx\approx 0.35-0.6~
\mathrm{GeV} \mathrm{fm}^{-1}$ at $T=500$~MeV, and for a jet energy of
$E=48\,T=24\,\mathrm{GeV}$.  Adjusting the color factors as appropriate
for SU(3) and extrapolating to the thermal density of gluons we
find $dE/dx\approx 2.5\pm 0.6$ \,GeV fm$^{-1}$.  At $T=400\,
\mathrm{MeV}$, the result is $dE/dx\approx 1.6\pm 0.4$ GeV fm$^{-1}$
($E=19.2\,\mathrm{GeV}$).

\begin{figure}[htb]
  \begin{center}
    \includegraphics[width=11cm]{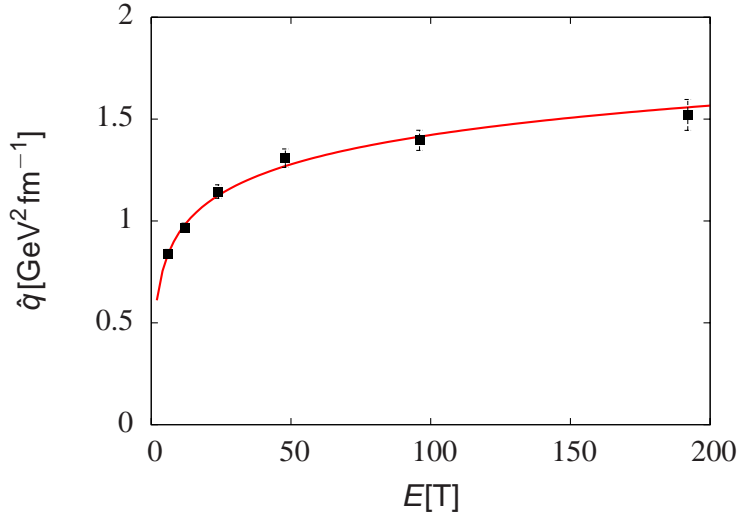}
    \caption{(Color online) Jet-energy dependence of $\hat{q}$ for $T=4$ GeV and
      $n=5$ fm$^{-3}$. $k^*\approx 1.16\,T$. The line shows
      the fit to Eq.~(\ref{<kt2>LL}), with an overall multiplicative
      factor of 0.63.}
    \label{fig:qhatE-T4-n5}
  \end{center}
\end{figure}

\begin{figure}[htb]
  \begin{center}
    \includegraphics[width=11cm]{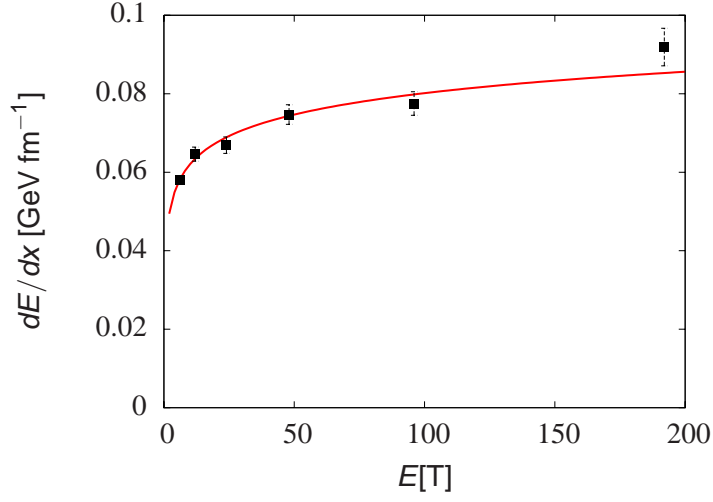}
    \caption{(Color online) Jet-energy dependence of $dE/dx$ for
      $T=4$~GeV and $n=5$~fm$^{-3}$. $k^*\approx 1.16\,T$. The line
      shows the fit to Eq.~(\ref{eq:dEdx}), with an overall
      multiplicative factor of 0.143.}
    \label{fig:dEdxE-T4-n5}
  \end{center}
\end{figure}

In Figs.~\ref{fig:qhatE-T4-n5} and~\ref{fig:dEdxE-T4-n5} we show how
$\hat{q}$ and $dE/dx$ depend on the energy $E$ of the jet. The
behavior is logarithmic, in agreement with the perturbative
expectation; compare to Eqs.~(\ref{<kt2>LL}) and (\ref{eq:dEdx}),
using $Q^2\sim ET$ (the explicit factor of $E$ in the numerator
cancels since $\langle s \rangle=6 E T$).  A fit of the numerical
result to (\ref{<kt2>LL}), using $Q^2\approx s$, leads to $C\approx
1.45$, but is good only if an additional prefactor of $\sim 0.63$ is
allowed. This suggests that the perturbative result does not describe
the numerical solution very well, which could perhaps be expected at
$g=2$. Repeating the analysis for $dE/dx$ via Eq.~(\ref{eq:dEdx}), we
find $C'\approx 7.7$; again, a multiplicative factor needs to be
included, this time it is $\sim 0.14$. Thus, for the jet energies
considered here, there is a smaller ``K factor'' relative to pQCD at
LL for $dE/dx$ than for $\hat{q}$; note that the former is sensitive
also to longitudinal momentum exchanges while the latter is not.

We have also determined the full $p_\perp^2$-distribution of the
high-momentum partons traversing the hot medium in order to assess the
relative contributions from various processes to its first moment
$\hat{q}$. We find that over time the initial $\delta$-function
broadens to a Gaussian distribution with a power-law tail. This
enhancement of transverse momentum broadening reflects the well known
result from QED and is in line with the findings for QCD in
Refs.~\cite{Gyulassy:2002yv,Qiu:2003pm}. The enhancement has also been
discussed within the higher twist formalism in
Ref.~\cite{Majumder:2007hx}. What is perhaps less obvious is the
relative magnitude of the Gaussian and power-law
parts, which may be expected to be time dependent. However, for time
scales typical of heavy-ion collisions we do not observe a large
relative shift of these contributions.

\begin{figure}[htb]
  \begin{center}
    \includegraphics[width=11cm]{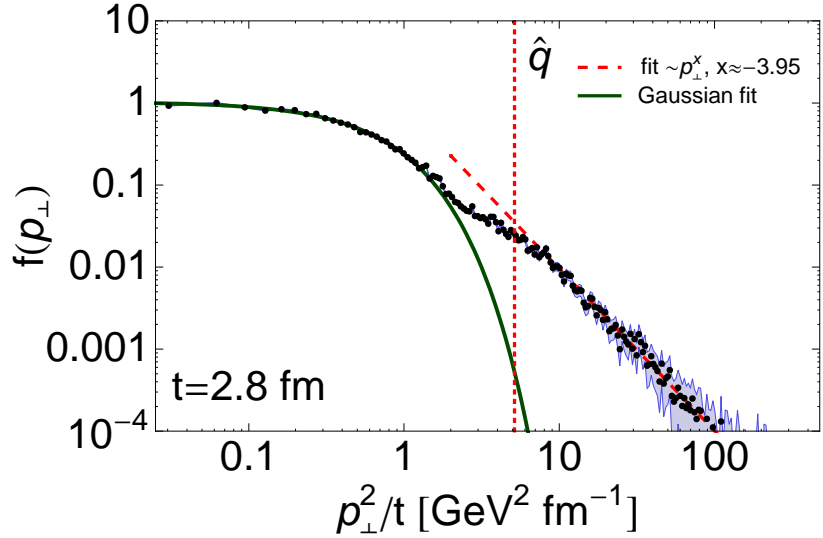}
    \caption{(Color online) $p_\perp^2$-distribution of the high-momentum ($E/T=48$)
      ``jet''-partons after $t\approx 2.8\,\mathrm{fm}$ for
      $T=4\,\mathrm{GeV}$ and $n=20\,\mathrm{fm}^{-3}$
      ($\hat{q}\approx 5.16\,\mathrm{GeV}^2\mathrm{fm}^{-1}$).}
    \label{fig:ptdistlogbin-n20-2.8fm}
  \end{center}
\end{figure}
\begin{figure}[htb]
  \begin{center}
    \includegraphics[width=11cm]{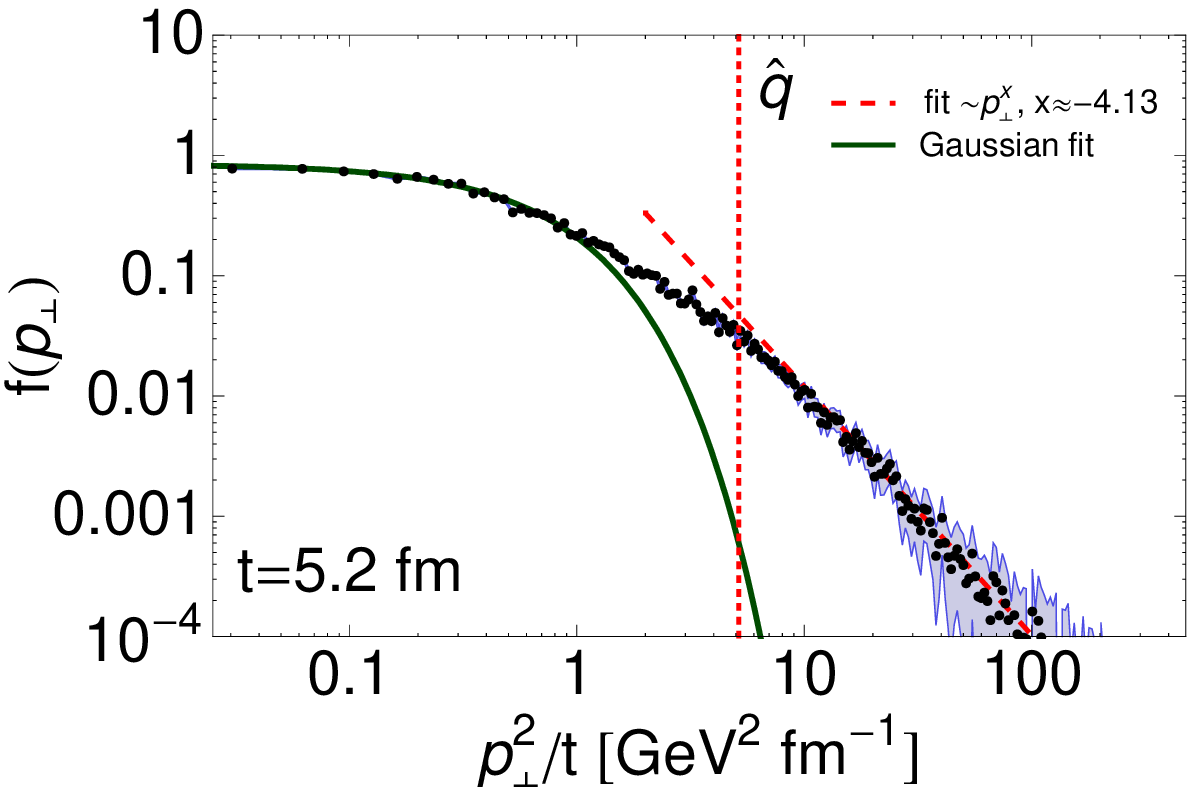}
    \caption{(Color online) $p_\perp^2$-distribution of the high-momentum ($E/T=48$)
      ``jet''-partons after $t\approx 5.2\,\mathrm{fm}$ for
      $T=4\,\mathrm{GeV}$ and $n=20\,\mathrm{fm}^{-3}$
      ($\hat{q}\approx 5.16\,\mathrm{GeV}^2\mathrm{fm}^{-1}$).}
    \label{fig:ptdistlogbin-n20-5.2fm}
  \end{center}
\end{figure}

Figs.~\ref{fig:ptdistlogbin-n20-2.8fm} and
\ref{fig:ptdistlogbin-n20-5.2fm} show the distribution of the
high-momentum ``jet'' test particles after $t\approx 2.8\,
\mathrm{fm}$ and $t\approx 5.2\, \mathrm{fm}$, respectively, in a
double-logarithmic plot versus $p_\perp^2/t$. We scale $p_\perp^2$ by
the inverse time so that the basic features of the distribution are
nearly time independent. 

The low-$p_\perp$ part follows a Gaussian distribution in $p_\perp$.
The power-law tail at large $p_\perp$ behaves approximately as $p_\perp^{-4}$.  This
is expected for particles experiencing only few scatterings since in
the high-energy limit the differential cross section
$d\sigma/dp_\perp^2 \sim p_\perp^{-4}$, c.f.~Eq. (\ref{eq:xsec}). This
is the probability distribution for the momentum transfer in a single
hard collision. In both figures we also indicate the value of
$\hat{q}$ to show that the power-law tail contributes significantly to
this transport coefficient. For the densities, temperatures and jet
energies considered here it is clearly not a very good approximation
to determine the transport coefficient $\hat{q}$ from the Gaussian part
of the distribution alone as this would underestimate $\hat{q}$
substantially: discarding the power-law tail from
Figs.~\ref{fig:ptdistlogbin-n20-2.8fm},\ref{fig:ptdistlogbin-n20-5.2fm}
gives $\hat{q}_{\text{Gauss}}\approx 0.6\,\mathrm{GeV}^2\,\mathrm{fm}^{-1}$.

Note also that transverse momenta on the order of the temperature
($T=4\,\mathrm{GeV}$), such as the separation scale $k^*\approx 1.2
T$, correspond to $p_\perp^2/t\approx 5.2\,
\mathrm{GeV}^2\mathrm{fm}^{-1}$ in
Fig.~\ref{fig:ptdistlogbin-n20-2.8fm} and to $p_\perp^2/t\approx 3\,
\mathrm{GeV}^2\mathrm{fm}^{-1}$ in
Fig.~\ref{fig:ptdistlogbin-n20-5.2fm}. Above this value for
$p_\perp^2/t$ the distribution is due almost entirely to hard
collisions (we have checked that multiple soft collisions do not
contribute much in that region).

Finally, we also provide an estimate for the nuclear modification
factor $R_{AA}$ of the jet spectrum due to elastic energy loss in a
classical Yang-Mills field\footnote{Classical radiative energy loss
  has recently been considered in ref.~\cite{Kharzeev:2008qr} but is
  not taken into account here.}. This is of relevance for collisions
of heavy nuclei at high energies: the large number of gluons produced
in the central rapidity region can be described as a classical field
for a short
time~\cite{Krasnitz:1998ns,Krasnitz:1999wc,Krasnitz:2001qu,Krasnitz:2002mn,Krasnitz:2003jw,Krasnitz:2002ng}
until the field modes decohere and
thermalize~\cite{Dumitru:2006pz}. These classical fields produced in
the early stage of the collision also exhibit long-range correlations
in rapidity~\cite{Dumitru:2008wn}, which we presently neglect.

We proceed as follows. From our simulations presented above, the elastic
energy loss at a density $n$ and for a separation scale $k^*\, (\equiv
\sqrt{3}\frac{\pi}{a}) = \sqrt{3}\,\, 2T$ can be parameterized as
\begin{equation}\label{eq:dEdx_cl}
\frac{dE}{dx}=K \, n \, \frac{16\pi\, \alpha_s^2 N_c^2}{N_c^2-1}\,
        \frac{1}{6T}\, \ln\left(C'^2\frac{6ET}{{k^*}^2}\right)\,,
\end{equation}
with $K=0.143$ and $C'=22.975$. This form for $dE/dx$ has been
established numerically in the weak-coupling regime specified by
Eq.~(\ref{eq:continuum}) for $N_c=2$; in what follows, we extrapolate
it to $N_c=3$ and to physical density and temperature. Note also that
for such large $k^*$ (on the order of the so-called ``saturation
momentum'' $Q_s$) most of the energy density is due to the classical
field. We evaluate this expression as a function of the jet energy for
$N_c=3$, $T=400$~MeV and the corresponding thermal density of
gluons. This corresponds to an energy density of about 17~GeV/fm$^3$,
which is an appropriate average over the first 1~fm/c of a central
Au+Au collision at RHIC
energy~\cite{Krasnitz:1998ns,Krasnitz:1999wc,Krasnitz:2001qu,Krasnitz:2002mn,Krasnitz:2003jw,Krasnitz:2002ng}.

\begin{figure}[htb]
  \begin{center}
    \includegraphics[width=11cm]{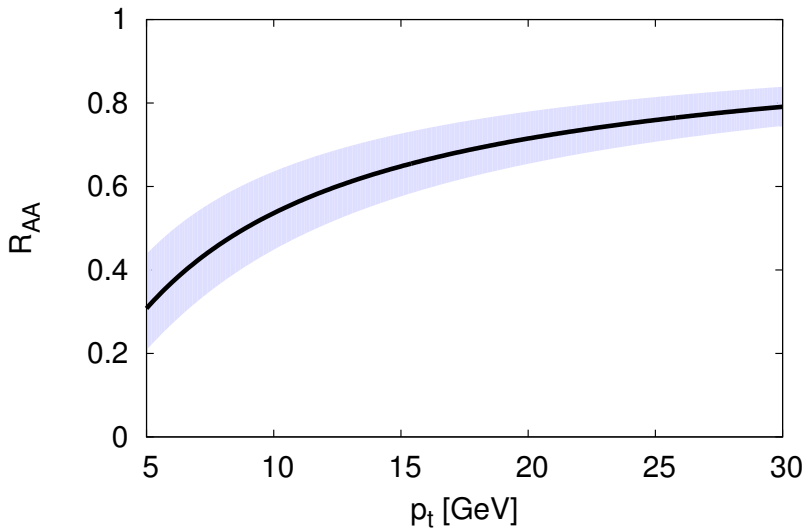}
    \caption{(Color online) Nuclear modification factor
      $R_{AA}(p_\perp)$ of jets due to elastic energy loss in a
      classical Yang-Mills field produced in the early stage of a
      relativistic heavy-ion collision at RHIC. The band indicates the
      uncertainty originating from the extrapolation of $dE/dx$ to
      physical temperatures (compare to Fig.~\ref{fig:dEdxT}).}
    \label{fig:R_AA}
  \end{center}
\end{figure}
The initial transverse momentum distribution of jets at RHIC can be
parameterized approximately as~\cite{Wicks:2005gt}
\begin{equation}
\frac{dN_i}{d^2p_\perp dy} \sim \frac{1}{p_\perp^{n+2}}
\end{equation}
with $n\approx 4$. Here, $p_\perp$ denotes the momentum of a jet
transverse to the colliding ion beams. In the central region
($y\sim0$) it is equal to the jet energy. The final distribution due
to interactions with the background is then given by
\bqa
\frac{dN_f}{d^2p_\perp dy} &=& \int d^2p_\perp' ~ \delta^{(2)}({\bf
  p_\perp} - (1-\epsilon) {\bf p_\perp'})\, \frac{dN_i}{d^2p_\perp' dy}
\nonumber\\
&= & \frac{1}{(1-\epsilon)^2} \frac{dN_i}{d^2p_\perp' dy}\Big|_{p_\perp' =
  \frac{p_\perp}{1-\epsilon} } = \frac{1}{p_\perp^{n+2}} \, (1-\epsilon)^n~.
\eqa
Here, $\epsilon$ denotes the fractional energy loss up to a time
$\tau$, which we take to be 1~fm/c:
\beq
\epsilon(p_\perp) = \frac{\tau}{p_\perp}\, \frac{dE}{dx}(p_\perp)~.
\eeq
Thus, the nuclear modification factor $R_{AA}$ at the parton level
(neglecting hadronization) can be written as~\cite{Wicks:2005gt}
\beq R_{AA}(p_\perp) = \frac{dN_f/d^2p_\perp dy}{dN_i/d^2p_\perp dy}
=\left( 1-\epsilon(p_\perp) \right)^n ~.  
\eeq 
We find that $\epsilon(p_\perp)$ is on the order of 10\% and that it
decreases with increasing jet energy. However, due to the relatively
steep initial spectrum of produced particles at RHIC, this can lead to
$\sim 30\%$ -- 50\% suppression in the $p_\perp$-range between 5~GeV
and 20~GeV; see Fig.~\ref{fig:R_AA}. Clearly, the experimentally
observed flat $R_{AA}\approx0.2$ can not be accounted for fully by
early-stage elastic energy loss in the classical field
background. Nevertheless, our result shows that this contribution is
significant and that it can not be neglected.



\section{Summary and Conclusions}
\label{sec:conc}
We have studied collisional energy loss as well as momentum broadening
of high-momentum gluon jets in a hot and dense non-Abelian SU(2)
plasma by solving the coupled system of Wong-Yang-Mills equations in
real time on a lattice. We separate the soft from the hard momentum
exchange interactions by introducing a separation scale $k^*$.  This
separation scale is given by the inverse lattice spacing $\sim 1/a$,
which determines the magnitude of the highest momentum field modes
that can be represented on the lattice. We fix its physical value to
be on the order of the temperature. Momentum exchanges below that
scale are mediated by the classical fields, those above the separation
scale by direct elastic collisions between the particles. The latter
were implemented through the pQCD collision kernel and the stochastic
method for determining scattering probabilities.

We restricted to collisional energy loss by simulating effectively
colorless jets (at the scale set by the lattice spacing). We were able
to obtain lattice-spacing and hence separation-scale independent
results for $\hat{q}$ and $dE/dx$ in a static and weakly coupled
plasma. The dependence on temperature and density, as well as on the
jet energy was found to follow, qualitatively, expectations from
pQCD. We then extrapolated our simulation results to thermal densities
and to more realistic temperatures which could not be simulated
directly. For a thermal gluon plasma (no quarks and anti-quarks) at
$T=400 \,\mathrm{MeV}$, and with color factors adjusted to the SU(3)
gauge group, we estimate $\hat{q}\approx 3.6\pm
0.3\,\mathrm{GeV}^{2}\mathrm{fm}^{-1}$ and $dE/dx\approx 1.6\pm 0.4\,
\mathrm{GeV} \mathrm{fm}^{-1}$, for a jet energy of
$19.2\,\mathrm{GeV}$. The errors are mainly due to the required
extrapolation. At finite time (on the order of the transverse
dimension of the collision zone in a heavy-ion collision), the
$p_\perp^2$-distribution of the high-momentum partons is found to be
well approximated by a Gaussian distribution at low
$p_\perp$ and to exhibit a power-law tail $\sim p_\perp^{-4}$ at high
$p_\perp$. The first moment of the distribution, i.e.\ the transport
coefficient $\hat{q}$, receives a large contribution from the
power-law tail.

We have also provided a first estimate of the (elastic) energy loss of
a jet traversing a classical Yang-Mills field, which might emerge in
the early stage of a collision of large nuclei at high energy. For a
field energy density of about 15~GeV/fm$^3$, the fractional energy
loss over a time interval of $\tau\simeq1$~fm/c amounts to about 10\%
-- 20\%. Once convoluted with the steep $\sim 1/p_\perp^6$ initial
spectrum of produced hard particles, this results in a nuclear
modification factor $R_{AA}\simeq0.5$ -- 0.8, indicating that energy
loss in the early stage (before the onset of hydrodynamic behavior of
the hot medium) may give a significant contribution to the observed
$R_{AA}$ at RHIC.

\section*{Acknowledgments}
We thank Oliver\ Fochler, Charles Gale, Sangyong Jeon, Berndt M\"uller, and Zhe\ Xu
for helpful discussions and comments. A.D.\ thanks
J.~Jalilian-Marian and D.~Kharzeev for emphasizing the importance of
energy loss in a classical Yang-Mills field.\\
The numerical simulations were performed
at the Center for Scientific Computing (CSC) of Goethe University,
Frankfurt am Main. M.S.\ and B.S.\ were in part supported by DFG Grant
GR 1536/6-1. B.S.\ gratefully acknowledges a Richard H.~Tomlinson
Fellowship awarded by McGill University as well as support from the
Natural Sciences and Engineering Research Council of Canada.  Y.N.\ is
supported by Japan MEXT grant No.~20540276. M.S.\ and Y.N.\
acknowledge support from the Yukawa Institute for Theoretical Physics
during the ``Entropy Production Before QGP'' workshop.

\bibliography{wymprc}

\end{document}